\documentclass[onecolumn,showpacs,preprintnumbers,amsmath,amssymb]{revtex4}
\usepackage[english]{babel}
\usepackage{graphics}
\usepackage{epsf}
\usepackage{bm}
\begin{document}

\title{Exact closed form analytical solutions for vibrating cavities}

\author{Pawe\l{} W\c{e}grzyn}
\affiliation{ Marian Smoluchowski Institute of Physics,
Jagellonian University, Reymonta 4, 30-059 Cracow, Poland}
\email{wegrzyn@th.if.uj.edu.pl}

\begin{abstract}

For one-dimensional vibrating cavity systems appearing in the
standard illustration of the dynamical Casimir effect, we propose
an approach to the construction of exact closed-form solutions. As
new results, we obtain solutions that are given for arbitrary
frequencies, amplitudes and time regions. In a broad range of
parameters, a vibrating cavity model exhibits the general property
of exponential instability. Marginal behavior of the system
manifests in a power-like growth of radiated energy.

\end{abstract}


\maketitle

\section{Introduction}

The dynamical variety of the Casimir effect has been studied in
numerous papers in the last two decades \cite{review, milton}. The
most favorable model for theoretical considerations refers to the
original Casimir's setup  \cite{casimir} of a cavity composed of
two perfectly conducting parallel plates. The dynamical
modification means that the distance between plates changes with
time. Then, we come across new fascinating phenomena in the
context of quantum field theory. The most impressive manifestation
of unusual non-classical properties of the theory of quantized
fields is the effect of particle production "from nothing". It is
referred as dynamical Casimir effect (DCE) or motion induced
radiation (MIR). Most of theoretical papers explore so-called
vibrating cavities, where oscillations of cavity walls are
periodic in time. Attention of their authors is attracted by the
instability of quantum fluctuations due to the parametric
resonance. They hope the resonance enhancement of particle
production could lead to macroscopic effects that would be
experimentally detected. Nowadays, the resonant instability of the
vacuum that is followed by an explosive particle production in a
vibrating cavity is believed to be accessible for  measurements
\cite{braggio,brown,kim,arbel,dodonov10}. However, we are still
awaiting for confirmation of successful experimental projects.

In this paper, we consider the behavior of the electromagnetic
field in a one-dimensional vibrating cavity. This problem was
studied initially by Moore \cite{moore}. A cavity with one
stationary wall and one moving wall with some prescribed
trajectory had been elaborated there. Moore did not solve any
particularly interesting cavity model, but he made a thorough
study of a general theory of solving such models. His basic
approach was generally used and developed  in next years
\cite{fulling}-\cite{haro}. From theoretical point of view, this
simplified model deals with several hard and important problems.
We mean the task of solving a wave equation with time-dependent
boundary conditions, the difficulties with analytical description
of physical systems under the parametric resonance conditions, the
problem of quantization of fields in limited regions with moving
boundaries, the squeezing of quantum states or the problems of
quantum entanglement and decoherence. One-dimensional vibrating
cavities provide the simplest theoretical laboratories to study
these issues in quantum field theory. The methods that are worked
out there can be essentially adapted for more complex models. Some
results and ideas are endorsed as well. There are advanced
approaches to include non-perfect \cite{schu2,schu3} or partly
transmitting \cite{haro} cavity walls, finite temperature effects
\cite{lambrecht2,schu5,jing} or proceed to three-dimensional case
\cite{dodonov, mundarain, dodonov6, dodonov7, dodonov8, crocce1,
crocce2, dodonov9}.

In the case of one-dimensional cavities and "scalar
electrodynamics", main achievements of numerous investigations
were obtained either in the framework of the effective Hamiltonian
approach \cite{law2,schu,review} or using  various numerical
approaches \cite{cole, wegrzyn1, li, fedotov, ruser1, ruser2,
ruser3}. Analytical solutions obtained through the perturbation
methods with effective Hamiltonians hold only for small amplitudes
\cite{dodonov3,dodonov5,dalvit1} or for particular time regimes,
either short time \cite{ji} or long time \cite{dodonov2} limits.
In many investigations, a frequency of cavity vibrations is
assumed to match a resonance frequency. It is well known  from
classical mechanics \cite{arnold} that parametric resonance occurs
also for frequencies that are not finely tuned provided that
respective amplitudes of oscillations are sufficiently large.  Off
resonant behavior of vibrating cavities is usually studied in the
limit of small detuning from resonance frequencies
\cite{dodonov4,wegrzyn2}.

Our aim is to gather  exact analytical and global solutions for
vibrating cavities.  In fact, there are few known solutions that
can be described by closed form expressions. The first closed form
exact solution to describe a cavity vibrating at its resonance
frequency was presented by Law \cite{law}. Law's solution
corresponds to a cavity that oscillates basically sinusoidally for
small amplitudes. The frequency of oscillations is twice the
lowest eigenfrequency of the cavity (so-called "principal
resonance" \cite{review}). Law found travelling wave packets in
the energy density of the field. He noted "sub-Casimir" quantum
fluctuations far away from the wave packets. Next, Wu et al.
\cite{wu} presented a family of exact analytical solutions for all
resonance frequencies. In the particular case of the second
resonance frequency, their solution is matching Law's solution.
They described emerging of wave packets in the energy  density,
indicated sub-Casimir fluctuations and emphasized the absence of
wave packets in the first resonant channel with the fundamental
cavity eigenfrequency ("semi-resonance" \cite{review}). One can be
puzzled that for any one of known exact solutions, there appears a
power-like resonance instability. Faithfully, the total energy of
the field increases quadratically with time there. On the other
hand, it is well recognized from other cavity models that the
total radiated energy typically grows exponentially with time
\cite{cole,meplan,dalvit1,wegrzyn1,petrov1,andreata}. For
instance, one can refer to the asymptotic formulas found by
Dodonov et al. \cite{review} for cavities that undergo harmonic
oscillations. It was generally argued  \cite{meplan,cole} that an
exponential resonant instability is typical for vibrating
cavities, while a power-like behavior constitutes  a critical
boundary between stability and instability regions defined by
domains of parameters \cite{wegrzyn3}. In this paper, we find
solutions that reveal  exponential instability generally and
exhibit a power-like law as a marginal effect.  Moreover, all
previously known exact solutions  applied only to resonance
frequencies. Here, we provide exact solutions that are adjustable
for all frequencies. This paper presents a rich class of exact and
closed-form solutions, in addition all formerly presented
solutions \cite{law,wu,review} are captured here as  particular
cases and examined in a more comprehensive way.

Our paper is organized as follows. In Section II, we present our
way of representing  solutions to describe the quantum dynamics in
a vibrating cavity. It relies on  $SL(2, R)$ symmetry of the
algebraic structure that exists for the quantized scalar field in
a static cavity \cite{gsw,wegrzyn4,wu2}. Actually, we abandon
Moore's function that is awkward in use. We put forward another
object, called a fundamental map by us, that has remarkable
analytical properties. These properties are collected in Section
III, together with appropriate mathematical formulae for primary
physical functions, namely the vacuum expectation of the energy
density of the field inside a cavity and the total radiated
energy. Finally, in Section IV we present a collection of exact
closed-form solutions. The results are summarized in Section V.

\section{Representation of solutions to describe the quantum dynamics in a vibrating cavity}

In the standard physical setup, we have an electromagnetic
resonator of length $L$ composed of two perfectly reflecting
walls. Initially, the cavity is static. Then, it undergoes
vibrations with a constant frequency $\omega$. In literature, it
is frequently assumed that the cavity length $L$ is related with
the period of oscillations $T=2\pi/\omega$. In this paper, we will
keep  the parameters $L$ and $T$ independent. The parameters
provide the characteristic physical length scales. The static
cavity length $L$ defines the magnitude of Casimir interactions.
In particular, it specifies the scale of quantum fluctuations
leading to the production of particles. The period $T$ is the
scale of parametric excitations of the system caused by some
external force. It is very useful in numerical computations to put
$T=\pi$. The parametric resonance is expected when $L$ and $T$ are
of the same order. Eventually, it depends also on an amplitude of
vibrations. We are willing to yield a phase diagram (Arnold's
diagram \cite{arnold}) that exhibits stability and instability
regions.

The derivation of the simplest mathematical model leads to the
quantization of free scalar field $A(x,t)$ with Dirichlet boundary
conditions imposed at the boundary walls $x=0$ and $x=L(t)$. The
trajectory of the oscillating wall is periodic: $L(t+T)=L(t)$. It
is important to assume that $L(t)>0$ (the cavity never collapses)
and $|\dot{L}(t)| \leq v_{max}<1$ (the wall velocity does not come
near the speed of light). Moreover, we impose $L(t)=L$ for $t<0$
(the cavity is static in the past, this condition is important for
the quantization). The construction of the basic set of solutions
for this problem was given in \cite{moore}:
\begin{equation}\label{base}
    A_N(t,x)=\frac{i}{\sqrt{4\pi N}}\left[ \exp{\left(-i\omega_N R(t+x) \right)}
     - \exp{\left(-i\omega_N R(t-x) \right)} \right] \ .
\end{equation}
The cavity eigenfrequencies $\omega_N=N\pi/L$ are called resonance
frequencies. We expect that  the parametric resonance occurs at
these frequencies for any amplitudes. However, the instability of
the system may appear also for other frequencies provided that the
amplitude of oscillations is sufficiently large. Usually, it is a
hard task to get the picture of the asymptotic behavior of the
system for any frequencies and amplitudes. Our knowledge of the
system comes to us through the Moore's function $R$ given by the
following equation:
\begin{equation}\label{moore}
    R(t+L(t))-R(t-L(t))=2L \ .
\end{equation}
Usually,  Moore's function $R$ is defined as a dimensionless
function (phase function). In this paper,  we will prefer to
define this function in dimensions of length. There is no general
theory of solving Eq.(\ref{moore}). Before we present a big set of
exact solutions of the above problem, it is worth to recall some
useful symmetry of the static cavity system \cite{wegrzyn4,wu2}.
In the static region for $t<0$, the quantized theory is invariant
under the conformal transformations:
\begin{equation}\label{conf}
    t \pm x \rightarrow R_{min}(t \pm x) \ ,
\end{equation}
with the functions $R_{min}$ defined by:
\begin{equation}\label{rmin}
    R_{min}(\tau)=\frac{2}{\omega_1}
    \arctan{\left( \sigma(\tan{\frac{\omega_1\tau}{2}}) \right)}\ ,
\end{equation}
where $\sigma(\tau)=(A\tau+B)/(C\tau+D)$ is any  homography and
$\omega_1$ is the lowest resonance frequency. Subsequent branches
of multivalued function \emph{arctan} should be always chosen and
linked together in such a way that a resulted function $R_{min}$
is continuous.   It is described here a well-known $SL(2,R)$
symmetry of free scalar fields quantized on a strip \cite{gsw}.
Surprisingly, this symmetry is rarely exploited in numerous papers
on physical models of the quantum field in a one-dimensional
cavity. In particular, the  symmetry helps to solve the puzzling
problem why there is no resonant behavior of the system for the
fundamental resonance frequency $\omega_1=\pi/L$.

In this paper, we will be searching for exact solutions of
Eq.(\ref{moore}) in the following form:
\begin{equation}\label{generalform}
    R(\tau)=\frac{2}{\omega}
    \arctan{\left( \Delta_{n(\tau)}(\tan{\frac{\omega\tau}{2}})
    \right)}\ + \ shift \ .
\end{equation}
In order to obtain closed-form solutions, we assume the range of
\emph{arctan} to be $[-\pi/2, \pi/2]$ (principal branch) and
appropriate shifts will be explicitly specified throughout. For
instance, the linear Moore's function, which describes a static
cavity, should be represented as:
\begin{equation}\label{linearfunction}
    R_{static}(\tau)=\tau - \frac{4\pi}{\omega}= \frac{2}{\omega}
    \arctan{(\tan{\frac{\omega\tau}{2}})}\ +
     \ \lfloor \frac{\omega\tau}{2\pi}-\frac{3}{2} \rfloor \, \frac{2\pi}{\omega} \
     ,
\end{equation}
where we have used the standard notation for the floor function.
The construction of the representation Eq.(\ref{generalform}) is
tied up with the well-known idea from classical mechanics
\cite{arnold}. To explore the dynamics of periodic systems with
parametric resonance, it is a handy way to deal with  mappings for
single periods. Here, we need a set of maps $\Delta_n$ numerated
by the number $n$. Fortunately, the maps are not independent. We
prove that it is enough to specify only the first map $\Delta_1$.
Henceforth, a function $\Delta_1(v)$ is going to be called a
fundamental map throughout this paper. This map defines the
auxiliary function $f$:
\begin{equation}\label{frep}
    f(\tau)=\frac{2}{\omega}
    \arctan{\left( \Delta_{1}(\tan{\frac{\omega\tau}{2}})
    \right)}\ + \ shift \ ,
\end{equation}
which is a solution of a simpler problem than Eq.(\ref{moore})
(see equations for billiard functions in \cite{wegrzyn3}):
\begin{equation}\label{f}
    f(t+L(t))=t-L(t) \ .
\end{equation}
Since the cavity is static in the past, we have always that
$f(\tau)=\tau-2L$ for $\tau<L$. The subject is also simplified due
to the fact that the auxiliary function $f$ fulfils the
periodicity condition:
\begin{equation}\label{periodicity}
    f(\tau+T)=f(\tau)+T \ .
\end{equation}
In general, the Moore's function $R(\tau)$ is not subject to any
periodic conditions. The reason lies in the lack of periodicity of
the index $n(\tau)$, that assigns a map to a particular point.

\begin{figure}[h]
\centering \leavevmode
$$\vbox{\epsfbox{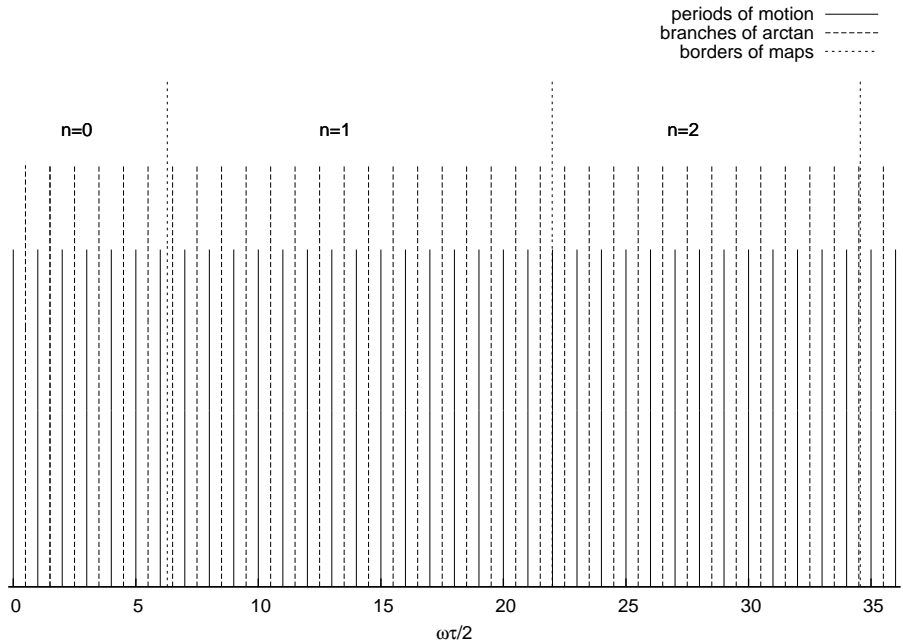} }$$
\caption{The borders of different change intervals for the Moore's
function R defined by Eq.(\ref{generalform}) with
Eq.(\ref{3d}).\label{fig1}}
\end{figure}

It is straightforward to prove that a fundamental map
$\Delta_1(v)$ designates unambiguously a Moore's function
$R(\tau)$. The solution of Moore's equation (\ref{moore}) can be
build according to the formula:
\begin{equation}\label{rf}
    R(\tau)=f^{\circ n(\tau)}(\tau) +2L[n(\tau)-1] \ .
\end{equation}
Looking at the representation Eq.(\ref{generalform}), one can
check easily: $\Delta_n=(\Delta_1)^{\circ n}$. Throughout this
paper, we use $(\Delta_{1})^{\circ n}$ to note $n$-fold
composition $\Delta_{1} \circ \Delta_{1} \circ ... \circ
\Delta_{1}$. It remains only to describe the step function
$n(\tau)$ that appears in Eq.(\ref{generalform}) and
Eq.(\ref{rf}).  As the function $f(\tau)$ is increasing, the
region for $\tau \geq L$ can be covered by intervals
$[L_{n-1},L_n)$, where $L_n \equiv (f^{-1})^{\circ n}(L)$.  The
map number $n(\tau)$ equals $n$ if the point $\tau$ lies inside
$[L_{n-1},L_n)$. Map markers $L_n$ will be called milestones
throughout this papers. If $\tau \in [L_{n-1},L_n)$, then $f(\tau)
\in [L_{n-2},L_{n-1})$. Thus, it is easy to find the following
recurrence relation, which is also very convenient for numerical
purposes:
\begin{equation}\label{n}
    n(\tau)=\left\{ \begin{array}{cc}
      0  & \ \ \ \tau<L \\
      1+n(f(\tau)) & \ \ \ \tau \geq L \\
    \end{array} \right.
\end{equation}
In order to provide a glimpse to details of future calculations
with the representation Eq.(\ref{generalform}) or Eq.(\ref{frep}),
we take a look at the solution given by Eq.(\ref{3d}). This
solution will be discussed later, but we glance over  the borders
of intervals for corresponding functions $f$ and $R$ there. They
are depicted in Fig.\ref{fig1}. Performing appropriate
calculations, one should take into account that all variables and
mappings are valid only in defined domains. Typically, the region
suitable for calculations of Bogoliubov coefficients or total
radiated energies is covered by two subsequent maps $\Delta_n$. It
makes evaluations of integrations and derivations of formulas more
complex. This is the price we have paid for the replacement of
Moore's equation by simpler relation Eq.(\ref{f}). However, we
will convince ourselves that this way is effective as a method for
obtaining analytical results. Later, the details of calculations
will be always skipped, so that the pattern in Fig.1 is the only
commentary on practical calculations.

It is difficult to derive the function $R(\tau)$ from
Eq.(\ref{moore}) for some prescribed trajectory $L(t)$. A great
number of numerical approaches and approximate solutions were
presented in other papers, but only few exact solutions are known.
One way to obtain exact solutions is to specify the function
$f(\tau)$, and then the trajectory $L(t)$ can be given in a
parametric form:
\begin{equation}\label{parametric}
    \left\{\begin{array}{ccc}
      t & = & [\tau+f(\tau)]/2 \\
      L(t) & = & [\tau-f(\tau)]/2 \\
    \end{array} \right.
\end{equation}
The prescribed function $f(\tau)$ represents an admissible
physical trajectory provide that it fulfils the following
requirements \cite{wegrzyn1}:
\begin{equation}\label{freq}
    \begin{array}{cl}
   (i) &   f(\tau)=\tau-2L \ \ \ {\rm for} \ \ \tau<L \\
    (ii) &  \frac{1-v_{max}}{1+v_{max}} \leq \dot{f}(\tau) \leq
       \frac{1+v_{max}}{1-v_{max}} \\
    (iii) &  f(\tau)<\tau \\
    \end{array}
\end{equation}
In this paper, we will be exploiting the representation
Eq.(\ref{generalform}) to describe solutions of equations for the
electromagnetic field in an oscillating one-dimensional cavity.
Before we start with the construction of solutions, we describe
general properties of  fundamental maps $\Delta_1$ extracted from
proper solutions.

\section{General properties of fundamental maps $\Delta_1$}
Knowledge of Moore's function enables us to draw out all
information about the vibrating cavity system.  The most important
object to calculate is the vacuum expectation value of the energy
density:
\begin{equation}\label{t00}
    \langle T_{00}(t,x)\rangle = \varrho(t+x)+\varrho(t-x) \ .
\end{equation}
Using appropriate formulas given in \cite{wegrzyn4} and our
representation Eq.(\ref{generalform}), we can easily calculate:
\begin{equation}\label{rho}
    \varrho(\tau)=-\frac{\omega^2}{48\pi}+\frac{\omega^2-\omega_1^2}{48\pi}
    \left[\frac{1+v^2}{1+\Delta_{n(\tau)}^2(v)}\Delta_{n(\tau)}'(v) \right]^2
    - \frac{\omega^2}{96\pi}(1+v^2)^2 S[\Delta_{n(\tau)}](v) \ ,
\end{equation}
where $v=\tan{(\omega\tau/2)}$ and $S[\Delta_{n(\tau)}](v)$
denotes the Schwartz derivative of $\Delta_{n(\tau)}$ with respect
to $v$. The total quantum energy radiated from the cavity can be
calculated from:
\begin{equation}\label{energy}
    E(t)=\int_0^{L(t)} \, dx \, \langle T_{00}(t,x)\rangle \, = \,
     \int_{t-L(t)}^{t+L(t)} \, d\tau \, \varrho(\tau) \, = \,
     \frac{2}{\omega} \, \int \, \frac{dv}{1+v^2} \, \varrho(v) \ .
\end{equation}
The most useful is the last formula which enables us to calculate
the total energy by integration with respect to $v$. However, we
should remember from the comment in the previous section on the
pattern in Fig.\ref{fig1} that the replacement
$v=\tan{(\omega\tau/2)}$ is valid only for a single period of
cavity motion. The interval of integration $[t-L(t),t+L(t)]$ is to
be divided into parts representing separate periods of motion. The
map number $n(\tau)$ may change at most once per period.

Let us remind the relation $\Delta_n=(\Delta_1)^{\circ n}$, and we
need only to specify the fundamental map $\Delta_1$. The knowledge
of this map makes it possible to predict the evolution of the
system and describe the resonance behavior. Henceforth, our
exploration of a quantum field theory system is quite similar to
examination of classical mechanics models under the parametric
resonance \cite{arnold}. We need only to analyze the asymptotic
behavior of iterations of the mapping ruled by $\Delta_1$, which
is known from the first period of motion.

We are going to make a list of general properties of fundamental
maps $\Delta_1$. First, we include information that the cavity is
assumed to be static in the past, i.e. for times $t<0$. It follows
that a solution for $t>0$ is to be sewed together with a static
one at $t=0$. In our context, there is no need to demand that the
sewing is perfectly smooth. For instance, we can accept that a
force which causes cavity motion may be suddenly switched on. Such
a solution may lead to  some Dirac delta terms in its function for
energy density, but from physical point of view the solution is
acceptable and useful for applications, so that it is definitely
worth saving them. Henceforth, let us propose some minimal set of
requirements for sewing. We put forward three sewing conditions at
the initial time $t=0$. The trajectory of the cavity wall and its
velocity should be continuous: $L(t=0)=L$ and $\dot{L}(t=0)=0$.
Moreover, there should be no sudden local growth of energy:
$\langle T_{00}(t=0,x)\rangle=-\pi/(24L^2)$, i.e. the local energy
density matches the Casimir energy density of vacuum fluctuations
at the initial time. It is now straightforward to gather a full
set of initial conditions for the fundamental map $\Delta_1(v)$:
\begin{equation}\label{initial}
    \begin{array}{ccc}
      \Delta_1(v_0) & = & - v_0 \\
      \Delta_1'(v_0) & = & 1 \\
      S[\Delta_1](v_0) & = & 0 \\
    \end{array} \ \ \ ; \ \ \ \ \ \ \
     \ \ \ v_0 \equiv \tan{\frac{\omega L}{2}}
     =\tan{\left(\pi \frac{L}{T}\right)}=\tan{\left(\frac{\pi}{2}
     \frac{\omega}{\omega_1}\right)} \ .
\end{equation}
The last condition implies that the construction of a fundamental
map is yet  a non-linear problem. Next, we impose the requirement
that the velocity of cavity wall should never exceed $v_{max}$.
The maximal velocity is a parameter of the cavity model and the
only limitation is that $v_{max}<1$. From Eq.(\ref{freq})(ii) we
obtain:
\begin{equation}\label{vmax}
\frac{1-v_{max}}{1+v_{max}} \leq
\frac{1+v^2}{1+\Delta_1^2(v)}\Delta_1'(v)
 \leq \frac{1+v_{max}}{1-v_{max}} \ .
\end{equation}
This is a strong constraint on possible maps. One immediate
consequence is that our function is increasing: $\Delta_1'(v)>0$.
We can also learn about its singularities from Eq.(\ref{vmax}). If
the function $\Delta_1(v)$ is  singular at some $v_s$:
\begin{equation}\label{limit1}
\lim_{v\rightarrow v_s \mp 0}\,\Delta_1(v)\, = \, \pm \, \infty \
,
\end{equation}
 then it is
easy to prove that the following limit is finite and different
from zero:
\begin{equation}\label{limit2}
  \lim_{v\rightarrow v_s}\,
\frac{\Delta_1'(v)}{\Delta_1^2(v)} \, = - \lim_{v\rightarrow v_s}
\, \frac{1}{(v-v_s)\Delta_1(v)} \ .
\end{equation}
Henceforth, it follows the function $\Delta_1(v)$ may have only
poles of order
 one:
 \begin{equation}\label{poles}
 \Delta_1(v)=\frac{h(v)}{(v-v_1)(v-v_2)...(v-v_s)} \ ,
\end{equation}
where the numerator $h(v)$ is an analytical function. Taking
Eq.(\ref{vmax}) together with Eq.(\ref{poles}), we note that  for
large values of $v$ the function $h(v)$ shows the following
asymptotic:
\begin{equation}\label{asymptotic}
    h(v) \ \sim \ |v|^k \ , \ \ \ \ \ \ k \, \in \, \{s-1,\ s, \ s+1\}
    \ .
\end{equation}
Finally, we look at the representation Eq.(\ref{frep}) and the
periodicity condition Eq.(\ref{periodicity}). We conclude that the
number of  singularities $s$ in the map $\Delta_1$ for the
representation Eq.(\ref{frep}) is at most one. Actually, we could
replace $\omega$ with $s\omega$ in  the representation
Eq.(\ref{frep}) and allow for more complex form defined by
Eq.(\ref{poles}). The same performance as that in Section IV might
give new exact closed form solutions, but we will not examine this
idea here.

In general, for large arguments either the function $\Delta_1$ is
unbounded or it takes a finite limit. Therefore,  the respective
continuity condition corresponds to one of two choices:
\begin{equation}\label{continuity}
\Delta_1(\pm \infty)=\pm \infty \quad \quad {\rm or} \quad \quad
\Delta_1(-\infty)=\Delta_1(+\infty)={\rm finite \  value} \ .
\end{equation}

Let us summarize this section. The basic set of solutions
Eq.(\ref{base}) for a quantum cavity system can be fully specified
by  Moore's function Eq.(\ref{generalform}). In turn, this
function is to be reconstructed from the fundamental map
$\Delta_1$. The fundamental map is associated with the first
period of motion. Some basic physical requirements lead to strong
mathematical conditions on the application of function $\Delta_1$
to cavity models which are admissible from physical point of view.
This includes suitable sewing conditions Eq.(\ref{initial}) at
some distinguished point $v_0$, the inequalities Eq.(\ref{vmax})
introduced by a limitation $v_{max}$ on a cavity wall velocity and
continuity condition Eq.(\ref{continuity}). Moreover, the function
$\Delta_1$ may have at most one singularity (only a simple pole)
and it behaves for large arguments according to
Eq.(\ref{asymptotic}) ($s$ is a number of singularities, i.e. 0 or
1 here).

For some given fundamental map $\Delta_1$ that fulfils all
required mathematical conditions, it may be still difficult to
derive  trajectory $L(t)$ from Eq.(\ref{frep}) and
Eq.(\ref{parametric}) or map ranges  $L_n$ and index function
$n(\tau)$ from Eq.(\ref{n}). However, it is possible for rational
functions. In the following section, we will discuss such
solutions. They form a big and interesting family of exactly
solvable cavity models. In particular, they include all examples
of exact closed form solutions on vibrating cavities known from
other papers \cite{law, wu}.

\section{Exact closed form analytical solutions}
We use the considerations of the previous sections to find a
family of exactly solvable quantum models of vibrating cavities.
The static cavity length $L$ is fixed and it characterizes a
physical scale. According to the naive understanding of parametric
resonance, the frequency of vibrations $\omega$ should be close to
one of the resonance frequencies $\omega_N$. It means that $L$ is
close to $N T/2$. However, it should be confirmed in a specific
cavity model whether this naive criterion of resonance is
justified. Moreover, it turns out  there is a more subtle
situation when $L$ is an odd multiplicity of $T/2$, i.e. the
parameter $v_0$ in Eq.(\ref{initial}) is infinite. Such cases
should be analyzed in our treatment separately.

\subsection{Linear fundamental maps $\Delta_1$}
We begin by addressing the case when a fundamental map $\Delta_1$
is a polynomial.  The condition Eq.(\ref{vmax}) that velocities
are not approaching the speed of light is very restrictive here.
It allows only for a linear function. First, we will  examine the
case when $v_0$ is finite.

\subsubsection{Finite values of $v_0$}

Our method of proceeding follows closely on the formalism
presented in the previous sections. Inserting a linear function
into conditions (\ref{initial}), we pick out:
\begin{equation}\label{linear}
\Delta_1(v)=v-2v_0 \ ; \ \ \  \ \ \ v \equiv
\tan{\frac{\omega\tau}{2}} \ .
\end{equation}
It is easy to verify that the above function fulfils all physical
requirements Eq.(\ref{initial}), Eq.(\ref{vmax}) and
Eq.(\ref{continuity}). Let us define the natural number $M$ and
the angle parameter $\theta$ by:
\begin{equation}\label{Mr}
    L=(M+\frac{\theta}{\pi})T \ \ ; \ \ \ \ \ \ \
     M=1,2,3,... \ , \  \ \ |\theta|<\frac{\pi}{2} \ .
\end{equation}
The parameter $M$ can be interpreted as the order of the
resonance. We will go through this subsection and see that the
parameter $M$ is better to characterize the resonance channel than
$N$.   The auxiliary function $f$ for $\tau \geq L$ from
Eq.(\ref{frep}) and its inverse function $f^{-1}$ for $\tau \geq
-L$ yield:
\begin{equation}\label{1f}
    \begin{array}{ccc}
      f(\tau) & = &  \frac{2}{\omega}
    \arctan{(\tan{\frac{\omega\tau}{2}-2v_0})}\ + \left(
    \lfloor \frac{\tau}{T} +\frac{1}{2} \rfloor -2M \right) T \ , \\
    & & \\
       f^{-1}(\tau) & = &  \frac{2}{\omega}
    \arctan{(\tan{\frac{\omega\tau}{2}+2v_0})}\ + \left(
    \lfloor \frac{\tau}{T} +\frac{1}{2} \rfloor +2M \right) T \ , \\
    & &  \\
     & & v_0=\tan{\frac{\omega L}{2}}=\tan{\theta} \ . \\
    \end{array}
\end{equation}
The corresponding trajectory of the cavity wall for $t\geq 0$ is
to be evaluated  from Eq.(\ref{parametric}). Using some
trigonometric identities, we reveal the following path:
\begin{equation}\label{1tr}
    L(t) = L+\frac{1}{\omega} \arcsin{(\sin{\theta} \cos{(\omega
    t)})}-\frac{\theta}{\omega} \  \ .
\end{equation}
\begin{figure}[h]
\centering \leavevmode
$$\vbox{\epsfbox{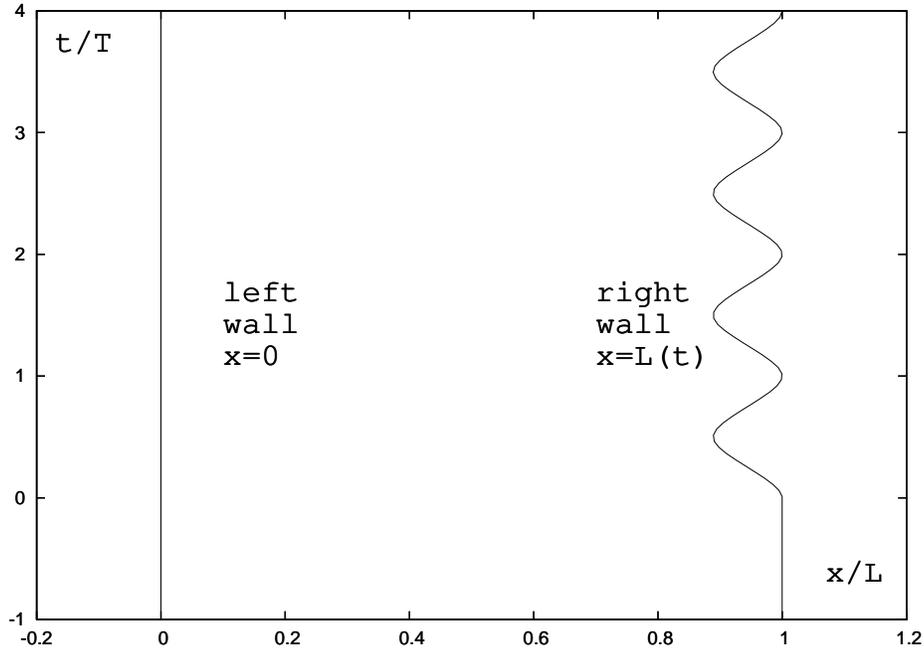} }$$
\caption{The trajectories of the cavity walls for the cavity
motion Eq.(\ref{1tr}) with $M=2$ and $\theta=\pi/4$.\label{fig2}}
\end{figure}
For small parameters $\theta$, the oscillations of the cavity wall
are close to a sinusoidal wave (see Fig.\ref{fig2}). With
increasing $\theta$, they are nearer to a triangle wave. The wall
oscillations  take place between $MT$ and $L$. The amplitude of
vibrations is $\Delta L=2|\theta|/\omega$, and the maximal
velocity yields:
\begin{equation}\label{1vmax}
    v_{max}=|\sin{\theta}| \ .
\end{equation}
The Moore's function for $\tau \geq L$ can be calculated from
Eq.(\ref{rf}):
\begin{equation}\label{1R}
    R(\tau)=\frac{2}{\omega}
    \arctan{\left(\tan{\frac{\omega\tau}{2}-2n(\tau)\tan{\theta}}\right)}\ +
   \ \  shift \ \ .
\end{equation}
\begin{figure}[h]
\centering \leavevmode
$$\vbox{\epsfbox{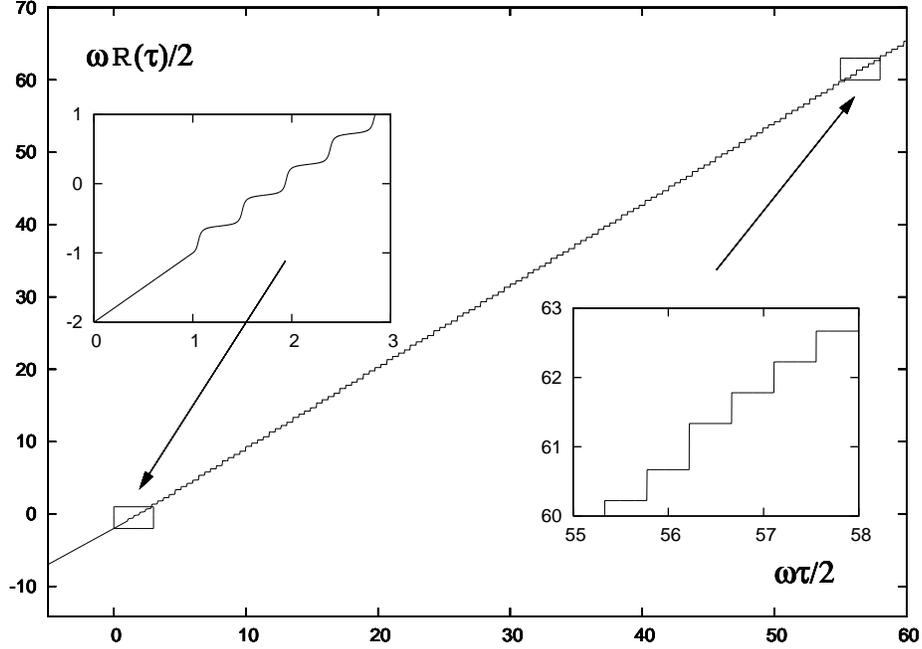} }$$
\caption{Moore's function for the cavity motion Eq.(\ref{1tr})
with $M=2$ and $\theta=\pi/4$.\label{fig3}}
\end{figure}
The representation Eq.(\ref{1R}) can be effectively used if we are
able to assign appropriate maps. It is nice that we are in a
position to calculate the milestones $L_n$ and the map number
$n(\tau)$ from Eq.(\ref{n}) exactly:
\begin{equation}\label{1n}
    \begin{array}{l}
      L_n=\frac{2}{\omega} \arctan{(2n+1)v_0}+(2n+1)M T \ , \\
      \\
      n(\tau)=n_0(\tau)-1+\Theta(\tau-L_{n_0(\tau)-1})+
      \Theta(\tau-L_{n_0(\tau)}) \ ,\\
      \\
      n_0(\tau)=\lfloor\tau /(2MT)+1/2\rfloor \ , \\
    \end{array}
\end{equation}
where the Heaviside step function is defined with $\Theta(0)=1$.
The Moore's function $R(\tau)$ from Eq.(\ref{1R}) for some
specific motion of type Eq.(\ref{1tr}) is shown in Fig.\ref{fig3}.
This function is always a small deviation from the linear function
Eq.(\ref{linearfunction}) that describes the static case.
Disturbances caused by the cavity motion are magnified in
Fig.\ref{fig3} for small and large function arguments. With
increasing arguments they approach a well-known staircase shape
("devil's staircase"). However, the steps are hardly regular. If
we look at the asymptotic behavior of the Moore's function, then
we are convinced that it is not a good object for practical
calculations, both analytical (perturbation methods) and
numerical. Thus, the transformations for phase functions like
Eq.(\ref{generalform}) are necessary to get a feasible way to
perform mathematical analysis of vibrating cavities in quantum
field theory.

The shape function for the energy density Eq.(\ref{rho}) for the
solution Eq.(\ref{linear}) reads:
\begin{equation}\label{1rho}
\rho(\tau)=-\frac{\omega^2}{48\pi}+\frac{\omega^2-\omega_1^2}{48\pi}
\left[ \frac{1+v^2}{1+(v-2 n(\tau) \tan{\theta})^2} \right]^2 \ .
\end{equation}
\begin{figure}[h]
\centering \leavevmode
$$\vbox{\epsfbox{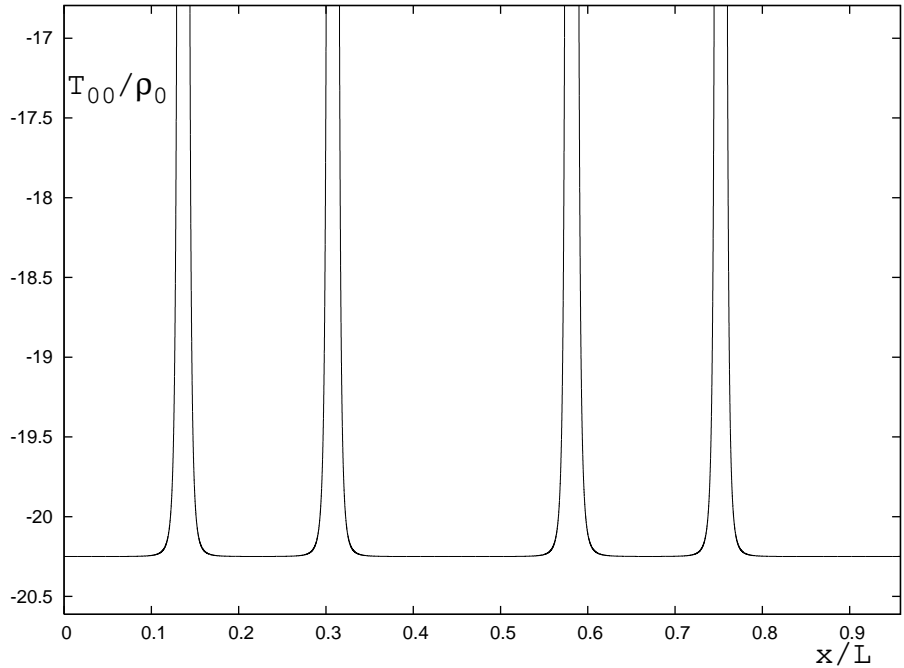} }$$
\caption{The energy density of the cavity Eq.(\ref{1tr}) with
$M=2$ and $\theta=\pi/4$ defined in terms of the Casimir energy
density of the static cavity $\rho_0=\pi/(24L^2)$.\label{fig4}}
\end{figure}
A snapshot of the energy density is displayed in Fig.\ref{fig4}.
In general, there are $M$ wave packets travelling left and $M$
wave packets travelling right. Their localization and their
evolution can be easily derived and it is in full agreement with
results of procedures described in \cite{wegrzyn3} and generalized
in \cite{wegrzyn5}. One can successfully derive periodic optical
paths and calculate cumulative Doppler factors, cumulative
conformal anomaly contributions and other quantities. Here, we
skip such details. Far from the narrow packets, in the so-called
sub-Casimir region \cite{law} the energy density is constant and
its asymptotic value is:
\begin{equation}\label{1asymp}
    T_{00}^{out}(\tau) \cong
    -\frac{\omega^2}{24\pi}=-\left(2M+\frac{2\theta}{\pi}
    \right)^2 \rho_0 \ \ , \ \ \ \ {\rm for \ large \ } \tau \ ,
\end{equation}
where $\rho_0=\pi/(24L^2)$ is the magnitude of Casimir energy
density for a static cavity of length $L$. Most of the energy is
concentrated in narrow wave packets. The heights of peaks are
proportional to $t^4$, and their widths shrinks like $t^{-2}$. It
suggests that the total energy grows with time like $t^2$. It is
true, and one can calculate from Eq.(\ref{energy}) an exact
formula. Here, we give only an asymptotic formula for large times:
\begin{equation}\label{1en}
E(t)\cong \frac{\omega(\omega^2-\omega^2_1)}{24 M \pi} \,
(\tan{\theta})^2 \, t^2 \ \ , \ \ \ \ t \gg 1 \ .
\end{equation}
As usual, there is no resonant behavior for the the lowest
resonance frequency. However, the resonance emerges for all
frequencies above this threshold: $\omega>\omega_1$ (or
equivalently: for $L>T/2$). Paradoxically, the resonance appears
here for all frequencies but resonance ones. For resonance
frequencies $\omega_N$, either the cavity is static or the motion
is singular (a triangular wave trajectory). Therefore, we should
learn that the resonance frequencies are auxiliary objects, and
real behavior of any physical cavity system depends on its
individual features. A specific feature of the solution
Eq.(\ref{1tr}) is that for some fixed initial cavity length $L$,
the resonance appear for almost all frequencies above some
threshold. However, there is no possibility to adjust the
amplitude of vibrations. There exists an exact relation between
the amplitude and the frequency:
\begin{equation}\label{1af}
    \frac{\Delta L}{L} = \left| 1-\frac{2\omega_1}{\omega} \,
    \lfloor \frac{\omega}{2\omega_1} +\frac{1}{2}\rfloor \right| \ .
\end{equation}
\begin{figure}[h]
\centering \leavevmode
$$\vbox{\epsfbox{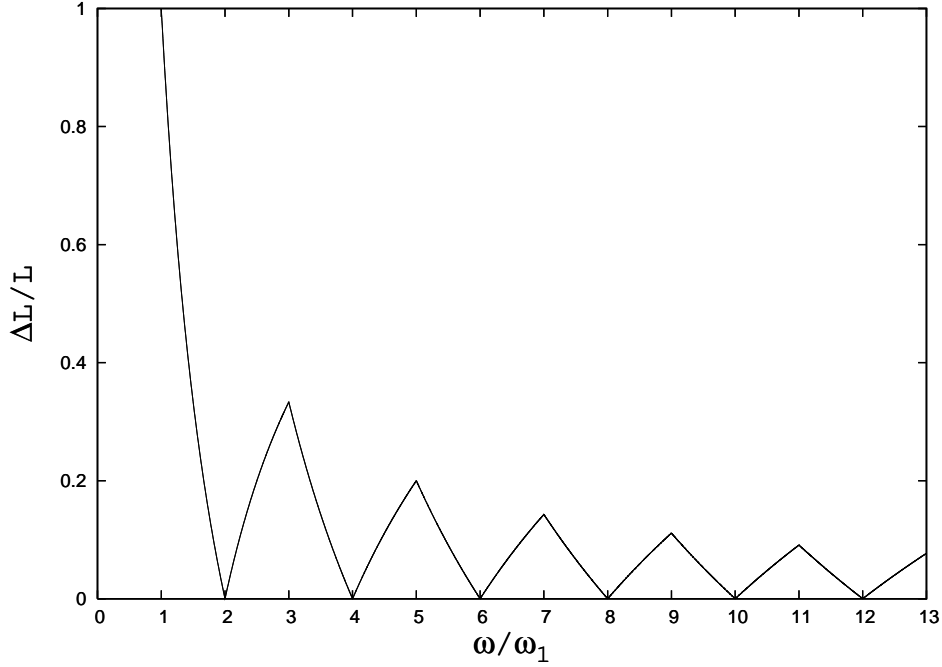} }$$
\caption{The phase diagram for the cavity model Eq.(\ref{1tr}):
the relative amplitude of cavity oscillations versus the frequency
as the multiplicity of the fundamental frequency.\label{fig5}}
\end{figure}
The most important problem for any linear dynamical system that
exhibits a parametric resonance phenomenon is to find stable and
unstable regimes for periodically excited parameters.  Usually,
the parametric resonance domains are depending on three crucial
parameters: frequency and amplitude of periodic excitation and
damping coefficient. The relevant Fig.\ref{fig5} exhibits the
phase diagram for the cavity model described in this section.
Since the frequency and the amplitude are related by
Eq.(\ref{1af}), the instable solutions Eq.(\ref{1tr}) are
represented by points on the curve to this plot. The instability
of solutions is quadratic according to Eq.(\ref{1en}). It was
suggested in \cite{wegrzyn3} that a cavity model with a power-like
instability appears as some boundary limit. If we extended the
model Eq.(\ref{1tr}) to possess more free parameters, then  other
points in Fig.\ref{fig5} would represent cavity motions when the
amplitude and the frequency do not match Eq.(\ref{1af}). Below the
border curve, the cavity system would be stable, and for states
represented by points that are placed above the curve we would
observe the resonance with the exponential growth of the total
radiated energy. The last example of solution considered in this
paper will justify such predictions. However, we are not able to
prove that the statement is generally true. Note famous Arnold's
tongue structure  \cite{arnold} in Fig.\ref{fig5}. However, the
tongues are rather broad. In the classical theories \cite{arnold},
Arnold's tongue has usually a narrow knife shape.

\subsubsection{Infinite values of $v_0$}
We get infinite values of sewing points in Eq.(\ref{initial}) if
the cavity oscillates at odd resonance frequencies
$\omega=\omega_{2M-1}$,
\begin{equation}\label{2Mr}
    L=(M-\frac{1}{2})T \ \ ; \ \ \ \ \ \ \
     M=1,2,3,...  \ .
\end{equation}
The conditions Eq.(\ref{initial}), Eq.(\ref{vmax}) and
Eq.(\ref{continuity}) are satisfied by a linear map with an
arbitrary intercept parameterized by $\theta$ (warning: $\theta$
has different meaning  that the same parameter in the previous
subsection. For convenience, we have redefined this parameter here
in such a way that numerous formulae match those of the previous
section):
\begin{equation}\label{2linear}
\Delta_1(v)=v-2\tan{\theta} \ ; \ \ \  \ \ |\theta|<\frac{\pi}{2}\
\  \ .
\end{equation}
The trajectory of the cavity wall $L(t)$ is the same as in
Eq.(\ref{1tr}). But now, the physical situation is different. In
the previous section, we were almost free to adjust the frequency
of the oscillations. If the frequency were fixed, the amplitude
would be given by Eq.(\ref{1af}). Here, the frequency is not
arbitrary, but the amplitude of the oscillations
$2|\theta|/\omega$ is adjustable. These solutions are already
known and they were presented first in \cite{wu} (they correspond
to the solutions numbered by $m\equiv 2N-1$ using the notation of
that paper). Our auxiliary functions are slightly modified:
\begin{equation}\label{2f}
    \begin{array}{ccc}
      f(\tau) & = &  \frac{2}{\omega}
    \arctan{(\tan{\frac{\omega\tau}{2}}+2\tan{\theta})}\ + \left(
    \lfloor \frac{\tau}{T} +\frac{1}{2} \rfloor -2M+1 \right) T \ , \\
    & & \\
       f^{-1}(\tau) & = &  \frac{2}{\omega}
    \arctan{(\tan{\frac{\omega\tau}{2}-2\tan{\theta}})}\ + \left(
    \lfloor \frac{\tau}{T} +\frac{1}{2} \rfloor +2M -1\right) T \ . \\
    \end{array}
\end{equation}
Again $v_{max}=|\sin{\theta}|$. The milestones $L_n$ and the map
number $n(\tau)$ are given by much simpler formulae:
\begin{equation}\label{2n}
    \begin{array}{l}
      L_n=(2n+1)L , \\
      \\
      n(\tau)=\lfloor \frac{\tau}{2L}+\frac{1}{2} \rfloor .\\
    \end{array}
\end{equation}
The Moore's function is given by the following formula:
\begin{equation}\label{2R}
    R(\tau)=\frac{2}{\omega}
    \arctan{\left(\tan{\frac{\omega\tau}{2}-2n(\tau)\tan{\theta}}\right)}\ +
   \lfloor\frac{\tau}{T}+\frac{1}{2}\rfloor T -2L \ .
\end{equation}
The profile function for the energy density is given by:
\begin{equation}\label{2rho}
\rho(\tau)=-\frac{(2M-1)^2\pi}{48L^2}+\frac{M(M-1)\pi}{12L^2}
\left[ \frac{1+v^2}{1+(v+2n(\tau)\tan{\theta})^2} \right]^2 \ .
\end{equation}
Now, it is much more easier to work out the integral
Eq.(\ref{energy}). Actually, it is interesting to look into an
exact and closed form formula for the total energy produced inside
the cavity:
\begin{equation}\label{2te}
    \begin{array}{l}
      E(t)=\frac{M(M-1)\pi\tan^2{\theta}}{12L^3} \, t^2+\\ \\
    +  \frac{M(M-1)\tan^2{\theta}}{3(2M-1)L^2}\left[ \frac{\pi}{2} \,
      {\rm sign}\, {\theta}-\arctan{\left( \frac{1}{\tan{(\omega
      t/2)}}\frac{1+\sqrt{1+\sin^2{\omega t}\tan^2{\theta}}-(t/L+1-2\alpha(t))\sin{\omega t}\tan{\theta}}
      {1-\sqrt{1+\sin^2{\omega t}\tan^2{\theta}+(t/L+1-2\alpha(t))\sin{\omega t}\tan{\theta}}}
      \right)}\right]\left( t+(1-2\alpha(t))L\right)+\\ \\
      -\frac{(2M-1)^2\pi}{24L^2}\,
      L(t)+\frac{M(M-1)\pi}{6L}+\frac{M(M-1)\pi
      \tan^2{\theta}}{3L}\alpha(t)(1-\alpha(t))+\\ \\
      \frac{M(M-1)\tan{\theta}}{3(2M-1)L} \, \frac{1+2(t/(2L)-\alpha(t))^2\tan^2{\theta}
      +(t/L+1-2\alpha(t))\tan{(\omega(t+L(t))/2)\tan{\theta}}}
      {1+(\tan{(\omega(t+L(t))/2)}+(t/L+2-2\alpha(t))\tan{\theta})^2} \ ,\\
    \end{array}
\end{equation}
where\begin{equation}\label{2alpha}
    \alpha(t) \equiv \frac{t}{2L} -\lfloor\frac{t}{2L}\rfloor \ .\\
\end{equation}
Similarly to the solution presented in the previous subsection,
the total energy of the system grows quadratically with time. We
have extracted the leading term. However, the next to leading
terms that are linear in time play an important role as well. They
cause that the energy is not irradiated continuously but rather in
sudden jumps. To verify that,  we should take two leading terms
from Eq.(\ref{2te}) and make the approximation for large values of
$t$. As a result we obtain:
\begin{equation}\label{1teap}
    E(t) \cong \frac{M(M-1)\pi \tan^2{\theta}}{3(2M-1)^2L} \, \left(
    \lfloor\frac{t}{T}\rfloor+\Theta(\theta)\right)^2 \,
\end{equation}
The presence of Heaviside function means that the problem is not
analytical in the parameter $\theta$, i.e. with respect to the
change of direction of oscillations. We see that impulses of
energy growth occur every period. For small amplitudes, the energy
is proportional to the square of the amplitude. This is in a
agreement with a non-relativistic limit of small velocities. It is
amusing to consider a quasi-classical analogue of the model.
Suppose, that at the initial state we have an uniform distribution
of energy of classical fields. The value of the energy density
equals the absolute value of the static Casimir energy:
$\rho_0=\pi/24L^2$. It corresponds to the classical potential
$A(t,x)=\varphi(t+x)+\varphi(t-x)$ with
$\varphi(\tau)=\pi\tau/48L^2$. The classical energy is given by
\cite{wegrzyn3}:
\begin{equation}\label{2ce}
E_{cl}t)=\int_0^{L(t)} \, dx \ T_{00}(t,x) \, = \,
     \int_{t-L(t)}^{t+L(t)} \, d\tau \ \dot{\varphi}^2(\tau)  \ .
\end{equation}
Next, we allow for the classical evolution of the electromagnetic
system. From classical equations of  motion we get:
$\varphi(\tau)=\varphi(f(\tau))$. Using initial conditions, we
obtain a classical global solution:
\begin{equation}\label{2cs}
    \varphi(\tau)=\frac{\pi}{48L^2} \ R(\tau) \ .
\end{equation}
We have encountered almost the same asymptotic formula for the
energy as in the quantum case. The only exception is the
coefficient:
\begin{equation}\label{1ceap}
    E_{cl}(t) \cong \frac{\pi \tan^2{\theta}}{12(2M-1)^2L} \, \left(
    \lfloor\frac{t}{T}\rfloor+\Theta(\theta)\right)^2 \,
\end{equation}
\begin{figure}[h]
\centering \leavevmode
$$\vbox{\epsfbox{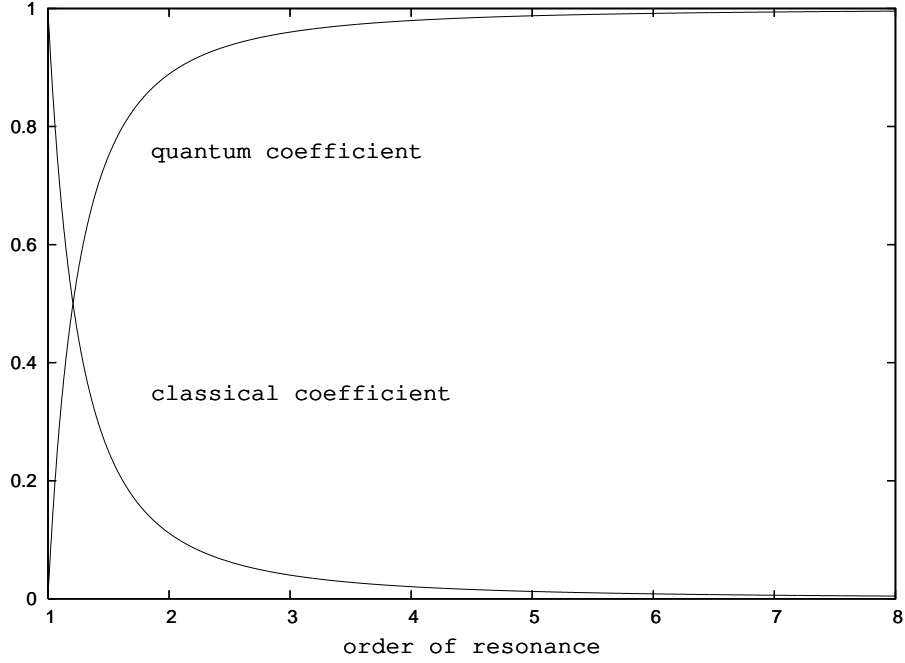} }$$
\caption{Coefficients for  quantum Eq.(\ref{1teap}) and classical
Eq.(\ref{1ceap}) asymptotic energy formulae. The coefficients are
in units of $\pi\tan^2{\theta}/12L$.
 \label{fig6}}
\end{figure}
The results are confronted in Fig.\ref{fig6}. We plot coefficients
of asymptotic energy formulae for the first eight resonance
channels. To make plots more readable, we have used continuous
lines. Classically, the strongest growth is for the fundamental
resonance frequency, next resonances are less effective. For the
quantum case, the situation reverses. Due to the effect of quantum
anomaly, there is no resonance in the first channel. Next, there
is a rapid saturation for higher resonance channels. In units of
$\pi \tan^2{\theta}/12L$, the sum of a classical coefficient and a
quantum coefficient is just a unity.

\subsection{Homographic fundamental maps $\Delta_1$ }

We now turn  the discussion to the case of maps $\Delta_1$ that
are rational functions with single poles:
\begin{equation}\label{deltasingular}
    \Delta_1(v)\, = \, \frac{h(v)}{v-v_0} \ .
\end{equation}
From Eq.(\ref{vmax}) we find that $h(v)$ is at most a quadratic
function. The periodicity condition Eq.(\ref{periodicity}) allows
only for one singularity of $\Delta_1$ per period. Therefore, we
can only consider homographic maps. It is convenient for us to
start the discussion with inversions, and then we will be looking
at a general case.

\subsubsection{Inversion map}
We evaluate that Eq.(\ref{initial}) and other necessary conditions
are satisfied by maps:
\begin{equation}\label{3d}
    \Delta_1(v)\, = \, -\frac{v_0^2}{v} \ , \ \ \ \ \
    v_0=\tan{\frac{\omega L}{2}}= \tan{\theta} \ .
\end{equation}
There are no solutions for singular $v_0$, so that $\omega \neq
\omega_{2N-1}$ and $|\theta|<\pi/2$. Moreover, we are forced to
assume $\theta \neq 0$ and this way all resonance frequencies are
excluded here: $\omega \neq \omega_N$. The auxiliary function $f$
for $\tau \geq L$ from Eq.(\ref{frep}) and its inverse function
$f^{-1}$ for $\tau \geq -L$ yield:
\begin{equation}\label{3f}
    \begin{array}{ccc}
      f(\tau) & = &  \frac{2}{\omega}
    \arctan{(v_0^2/\tan{\frac{\omega\tau}{2}})}\ + \left(
    \lfloor \frac{\tau}{T}  \rfloor -2M+\Theta(-\theta) \right) T \ , \\
    & & \\
       f^{-1}(\tau) & = &  \frac{2}{\omega}
    \arctan{(v_0^2/\tan{\frac{\omega\tau}{2}})}\ + \left(
    \lfloor \frac{\tau}{T} \rfloor +2M +\Theta(\theta)\right) T \ , \\
    & &  \\
     & & L=(M+\frac{\theta}{\pi})T \ . \\
    \end{array}
\end{equation}
The trajectory of the cavity wall for $t>0$ is reconstructed from
Eq.(\ref{parametric}):
\begin{equation}\label{3tr}
    L(t) = L-\frac{2\theta}{\omega}+\frac{{\rm sign} \, \theta}{\omega}
    \left[
    \frac{\pi}{2}-\arcsin{(\cos{2\theta} \cos{(\omega t)})}
    \right] \  \ .
\end{equation}
Evidently, the maximal velocity is now:
\begin{equation}\label{3vmax}
    v_{max}=|\cos{2\theta}| \ .
\end{equation}
In the limit $\omega \rightarrow \omega_{N}$, we encounter a
triangle wave trajectory.  For $\omega =(\omega_N+
\omega_{N-1})/2$, our solution degenerates to a static one. The
oscillations do always take place between $L$ and $L+{\rm
sign}\theta(\pi-4|\theta|)/\omega$. The corresponding milestones
for our representation of Moore's function are given by:
\begin{equation}\label{3mile}
    L_n=(-1)^n L+\lfloor \frac{n+1}{2}\rfloor (4M+{\rm sign}
    \, \theta) T \ .
\end{equation}
We make the energy density explicit:
\begin{equation}\label{3enden}
    \rho(\tau)= \left\{ \begin{array}{cc}
      -\frac{\omega_1^2}{48\pi} & {\rm for} \ \ \tau \in [L_{2k-1},L_{2k})\\
      & \\
      -\frac{\omega^2}{48\pi}+
      \frac{\omega^2-\omega_1^2}{48\pi}
 \frac{v_0^2(1+v^2)}{v_0^4+v^2}
      & {\rm for} \ \ \tau \in [L_{2k},L_{2k+1})\\
    \end{array} \right.
\end{equation}
There are wave packets in the energy density, but there is no
unbounded growth of the total energy. The quantum cavity system is
stable and its total accumulated energy oscillates with the period
$(4M+{\rm sign} \, \theta)T$.

The solution is also well-defined for $\omega<\omega_1$ ($L<T/2$).
It corresponds to $M=0$ and $\theta>0$ in Eq.(\ref{3f}) and
Eq.(\ref{3tr}). Here, as the only effect of the cavity motion
there are pits in the energy density (negative wave packets) that
may appear periodically in synchronization with cavity
oscillations.

\subsubsection{Homographic map}

As a final and the most interesting application of our ideas, we
consider a solution with a fundamental map being a homographic
function. So then, upon confrontation with initial conditions
Eq.(\ref{initial}), we set:
\begin{equation}\label{4d}
    \Delta_1(v)\, = \, -\frac{v_1 v+v_0(v_0-2v_1)}{v-v_1} \ ,
\end{equation}
where $v_0=\tan{(\omega L/2)}$ and $v_1$ is an arbitrary
parameter. It is straightforward to check that physical solutions
exist on condition that $v_0 \neq v_1$.  In passing, we note that
the solutions that have been described in the previous subsection
are reproduced for $v_1=0$.

The evaluation of  relevant auxiliary functions $f$ and $f^{-1}$
ends with the results:
\begin{equation}\label{4f}
    \begin{array}{ccc}
      f(\tau) & = &  \frac{2}{\omega}
    \arctan{\Delta_1(v)}\ + \left(
    \lfloor \frac{\tau}{T} +\frac{1}{2}
      \rfloor -2M+\Theta(v-v_1)-\Theta(v_0-v_1) \right) T \ , \\
    & & \\
       f^{-1}(\tau) & = &  \frac{2}{\omega}
    \arctan{\Delta_1^{-1}(v)}\ + \left(
    \lfloor \frac{\tau}{T} +\frac{1}{2}
    \rfloor +2M +\Theta(v+v_1)-\Theta(v_1-v_0)\right) T \ , \\
    & &  \\
     & & L=(M+\frac{1}{\pi}\arctan{v_0})T \ . \\
    \end{array}
\end{equation}
The milestones are given by:
\begin{equation}\label{4mile}
    L_n=\frac{2}{\omega} \arctan{\Delta^{-1}_{n}(v_0)}+\left[
    (2n+1)M+\sum_{k=0}^{n-1} \Theta(\Delta^{-1}_{k}(v_0)+v_1)
    -n \Theta(v_1-v_0) \right] T \ .
\end{equation}
The angle parameter $\theta$ may be introduced here by using the
following formula:\begin{equation}\label{4theta}
    \tan{\theta}=\frac{1+v_0^2}{2v_1}-v_0 \ .
\end{equation}
With the above definition, the derivation of the trajectory of the
cavity wall from Eq.(\ref{parametric}) gives us:
\begin{equation}\label{4trpre}
    \sin{(\omega L(t)+\theta)}=\sin{(\omega L+\theta)}\cos{\omega
    t} \ ,
\end{equation}
and it can be disentangled successfully:
\begin{equation}\label{4tr}
    L(t) = L+\frac{1}{\omega}\left[\arcsin{(\sin{(\omega L+\theta)}
    \cos{(\omega t)})}
    -\arcsin{(\sin{(\omega L+\theta)})} \right] \  \ .
\end{equation}
We have assumed throughout this paper that the functions $arcsin$
and $arctan$ have their ranges restricted to $[-\pi/2,\pi/2]$. It
makes the right hand side of Eq.(\ref{4tr}) uniquely and properly
defined. The maximal velocity is:
\begin{equation}\label{4vmax}
    v_{max}=|\sin{(\omega L+\theta)}| \ ,
\end{equation}
while the amplitude of oscillations is given by $\Delta
L=(2/\omega) \arctan{v_{max}}$.

It is important to point some special cases of Eq.(\ref{4tr}). For
$M=1$, $v_0=0$ and $v_1=1/(2\tan{\theta})$, we get a cavity model
investigated by Law in \cite{law}. It was the first exact closed
form solution presented in literature. The generalization of this
solution for any $M$ is leading to a second set of solutions
described in the paper by Wu et al. \cite{wu} (the solutions with
$m=2N$ in their notation). All solutions correspond to cavity
vibrations with resonance frequencies $\omega=\omega_N$. It was
established for Law's and Wu's solutions that there appears
resonant instability with a power-like behavior. We are not going
to discuss these solutions here and send the reader back to the
original papers. We wish only to note that there is one more class
of solutions with a power-like instability. We obtain these
solutions if we put $v_0=2v_1$ in Eq.(\ref{4d}). Let us describe
them very briefly. The frequency of oscillation is a free
parameter and may be tuned to any value greater than the
fundamental frequency $\omega_1$. But the amplitude of
oscillations is uniquely defined by the choice of frequency.

The solutions with exponential instability are obtained for $v_0
\neq 0$ and $v_0 \neq 2v_1$. The derivation of the maps $\Delta_n$
for the representation of Moore's function Eq.(\ref{generalform})
requires the calculation of $n$-fold composition of the
fundamental map given by Eq.(\ref{4d}). It is easy for
homographies, so that the result is:
\begin{equation}\label{4dn}
    \Delta_n(v)=\Delta_1^{\circ n}(v)=
    (\lambda_1-\lambda_2)\frac{2(\lambda_1^n+\lambda_2^n)v+(\lambda_1-\lambda_2)
    (\lambda_1^n-\lambda_2^n)}{4(\lambda_1^n-\lambda_2^n)v
    +2(\lambda_1-\lambda_2)(\lambda_1^n+\lambda_2^n)} \ ,
\end{equation}
where:
\begin{equation}\label{4lambda}
    \lambda_{1,2}=-v_1 \pm \sqrt{v_0(2v_1-v_0)} \ .
\end{equation}
The profile function for the energy density is then given by:
\begin{equation}\label{4rho}
\rho(\tau)=-\frac{\omega^2}{48\pi}+\frac{\omega^2-\omega^2_1}{48\pi}
(v_0-v_1)^{2n(\tau)} \left[
\frac{1+v^2}{(\frac{\lambda_1^n+\lambda_2^n}{2}v+\frac{(\lambda_1-\lambda_2)
(\lambda_1^n-\lambda_2^n)}{4})^2+(\frac{\lambda_1^n-\lambda_2^n}{\lambda_1-\lambda_2}v
    +\frac{\lambda_1^n+\lambda_2^n}{2})^2}
\right]^2 \ .
\end{equation}
We restrict ourselves to calculate only the approximate value of
the total energy for large times. Therefore, the first term in
Eq.(\ref{4rho}) can be omitted, while the second term integrated
over one period gives:
\begin{equation}\label{rach1}
 \int_{v=-\infty}^{v=+\infty} \, d\tau \, \rho(\tau)=
\frac{\omega^2-\omega^2_1}{48\pi} \, \frac{Tr(H^TH)}{det(H)} \ ,
\end{equation}
where $H$ is a matrix composed of coefficients of the homography
$\Delta_n$ in Eq.(\ref{4dn}). It can be easily calculated that:
\begin{equation}\label{rach2}
 \frac{Tr(H^TH)}{det(H)}=\frac{1}{4}\left(v_1+\frac{1}{v_1}\right)^2\left[
 \left(\frac{\lambda_1}{\lambda_2}\right)^{n(\tau)}
 + \left(\frac{\lambda_2}{\lambda_1}\right)^{n(\tau)}
 \right]-\frac{1}{2}\left(v_1-\frac{1}{v_1}\right)^2 \ .
\end{equation}
For simplicity, we have assumed here that the number $n(\tau)$ do
not change in the interval of integration. Further, we assume
small amplitudes: $2L(t)\approx2L\approx2MT$ and obtain the
approximate formula:
\begin{equation}\label{4enapr}
    E(t) \cong \frac{4M^2-1}{96}\left(v_1+\frac{1}{v_1}\right)^2
    \cosh{\left[  \frac{v_1-
    \sqrt{v_0(2v_1-v_0)}}{v_1+\sqrt{v_0(2v_1-v_0)}}
    \, \frac{t}{2L}\right]} \ .
\end{equation}
In the above brief calculation, we have demonstrated that it is
rather easy and safely in our treatment to perform approximate
calculations and skip insignificant details. In fact, the
treatment described in Section II and III is well adapted for
perturbative methods. However, the relation for the minimal
amplitude of oscillations $\Delta L_{min}$ enough to trigger
exponential instability of the cavity system vibrating at some
fixed frequency $\omega$ can be derived exactly:
\begin{equation}\label{4af}
    \frac{\Delta L_{min}}{L} = \frac{\left| \frac{\omega}{\omega_1}
    - \lfloor \frac{\omega}{\omega_1} +\frac{1}{2}\rfloor \right|}{
    \frac{\omega}{\omega_1} } \ .
\end{equation}
\begin{figure}[h]
\centering \leavevmode
$$\vbox{\epsfbox{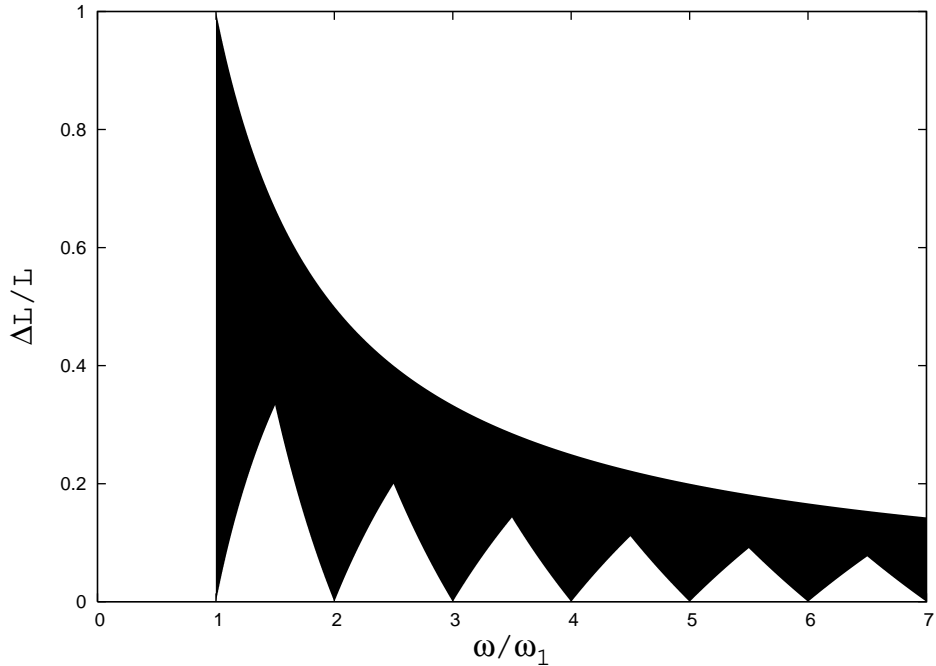} }$$
\caption{The phase diagram for stability and instability regions
for the cavity model Eq.(\ref{3tr}).\label{fig7}}
\end{figure}
As the velocity of cavity wall cannot reach a speed of light, we
obtain also the upper limit for amplitudes of oscillations $\Delta
L_{min}$ at given frequency $\omega$:
\begin{equation}\label{4af2}
    \frac{\Delta L_{max}}{L} =  \frac{\omega_1}{\omega}
     \ .
\end{equation}
 The above relation allows us to set up the phase diagram of
stability and instability regions. A black area in Fig.\ref{fig7}
correspond to the instability region of the vibrating cavity model
Eq.(\ref{3tr}). Below this area, we have defined frequencies and
amplitudes the cavity model is stable. A marginal behavior appears
when the energy grows quadratically with time. It is observed for
$v_0=0$ (resonance frequencies, boundaries between adjacent
Arnold's tongues) and $v_0=2v_1$ (boundary points between
stability and instability regions). Other parts of the diagram
correspond to points where the cavity model is not well defined
(physical assumptions about $L(t)$ at the beginning of Section II
are violated).

\section{Conclusions}

We have presented a rich class of exact and closed form analytical
solutions for the quantum vacuum field in a one-dimensional cavity
vibrating under the parametric resonance conditions. The solutions
are valid for all times,  frequencies of cavity oscillations
and/or their amplitudes are free  parameters. For small
amplitudes, cavity oscillations are close to sinusoidal ones. In
view of these properties, we can expect our solutions to yield all
generic features known from other investigations on vibrating
cavity models in a single dimension.

The representation of solutions Eq.(\ref{generalform}) that
appears in our treatment is based on $SL(2,R)$ symmetry of scalar
fields quantized in a static cavity. We have introduced the notion
of fundamental maps that are more convenient to proceed  than
Moore's phase functions. There is a direct mathematical
relationship between iterations of fundamental maps and the
mechanism of parametric resonance. This is the way we can get
insight into the regions of stability and instability of the model
(see Fig.\ref{fig7}). One can calculate the rate of increase of
the energy and the Lyapunov exponents. The stability-instability
transition points and points between adjacent Arnold's tongues
correspond to cavity models with a power-like instability. Thus,
the most crucial questions can be tackled. If we insist on
detailed calculations or exact formulas, then we have to set up
how regions in space are covered by our maps (see Fig.\ref{fig1}).
Summarizing technical aspects, we can tackle with any solution
successfully and completely provided that we know its fundamental
map $\Delta_1$ Eq.(\ref{frep}) and respective ranges of maps $L_n$
together with $n(\tau)$ Eq.(\ref{n}). In this paper, general
properties of fundamental maps for any physically reasonable
solutions have been described. This setup is also a good start
point for perturbative calculations.

To the best of our knowledge, both exact closed form  solutions
for off resonant frequencies and exact closed form  solutions with
exponential instability for vibrating cavities were not presented
before. It refers also to Arnold's phase diagrams of
stability-instability regions for solutions of vibrating cavity
models. It is also important to state that the mechanism of
parametric resonance in the quantum field theory shares many
common features with its analogue in the classical theory. We
believe that this similarity could be maintained also for
three-dimensional vibrating cavities. The same should be true for
the relevance of the symmetry of the quantized cavity model.

\end{document}